\documentclass[12pt]{article}
\usepackage{amsfonts, amsmath, amsthm, amssymb, graphicx, latexsym, slashbox,natbib, color, bm}

\newcommand{\blind}{0}

\begin{document}

\def\spacingset#1{\renewcommand{\baselinestretch}%
{#1}\small\normalsize} \spacingset{1}

\if0\blind
{\date{ }
  \title{\bf Sparse canonical correlation analysis from a predictive point of view}
  \author{Ines Wilms\thanks{
    Financial support from the FWO (Research Foundation Flanders) is gratefully acknowledged (FWO, contract number 11N9913N).}\hspace{.2cm}\\
    Faculty of Economics and Business, KU Leuven\\
    and \\
    Christophe Croux \\
    Faculty of Economics and Business, KU Leuven}
  \maketitle
} \fi

\if1\blind
{
  \bigskip
  \bigskip
  \bigskip
  \begin{center}
    {\LARGE\bf Sparse canonical correlation analysis from a predictive point of view}
\end{center}
  \medskip W
} \fi

\bigskip
\begin{abstract}
Canonical correlation analysis (CCA) describes the associations between two sets of variables by maximizing the correlation between linear combinations of the variables in each data set. However, in high-dimensional settings where the number of variables exceeds the sample size or when the variables are highly correlated, traditional CCA is no longer appropriate. This paper proposes a method for sparse CCA. Sparse estimation produces linear combinations of only a subset of variables from each data set, thereby increasing the interpretability of the canonical variates. We consider the CCA problem from a predictive point of view and recast it into a regression framework. By combining an alternating regression approach together with a lasso penalty, we induce sparsity in the canonical vectors.  We compare the performance with other sparse CCA techniques in different simulation settings and illustrate its usefulness on a genomic data set. 
\end{abstract}

\noindent%
{\it Keywords:}  Canonical correlation analysis; Genomic data; Lasso; Penalized regression; Sparsity.

\spacingset{1.45}
\newpage
\section{Introduction}
The aim of canonical correlation analysis (CCA), introduced by \cite{Hotelling36}, is to identify and quantify linear relations between two sets of variables. CCA is used in various research fields to study associations, for example, in physical data \citep{Pison04}, biomedical data \citep{Kustra06}, or environmental data \citep{Iaci10}. One searches for the linear combinations of each of the two sets of variables having maximal correlation. These linear combinations are called the \textit{canonical variates} and the correlations between the canonical variates are called the \textit{canonical correlations}. We refer to e.g. \citeauthor{Johnson98} (\citeyear{Johnson98}, Chapter 10) for more information on canonical correlation analysis.

At the same time, we want to induce sparsity in the canonical vectors such that the linear combinations only include a \textit{subset} of the variables. Sparsity is especially helpful in analyzing associations between high-dimensional data sets, which are commonplace today in, for example, genetics \citep{Qi14} and machine learning \citep{Sun2011, Liu14}. %, and economics \citep{Fan2011}
Therefore, we propose a sparse version of CCA where some elements of the canonical vectors are estimated as exactly zero, which eases interpretation. For this aim, we use the formulation of CCA as a prediction problem.

Consider two random vectors ${\bf x} \in \mathbb{R}^{p}$ and ${\bf y} \in \mathbb{R}^{q}$. We assume, without loss of generality, that all variables are mean centered and that $p \leq q$. Denote the joint covariance matrix of (${ \bf x }$,${ \bf y }$) by
\begin{equation}
{ \bf \Sigma } =  \begin{bmatrix}  { \bf \Sigma_{xx} } & { \bf \Sigma_{xy} }  \\ { \bf \Sigma_{yx} } & { \bf \Sigma_{yy} } \end{bmatrix} \label{covmatrix}
\end{equation}
with $r=rank({ \bf \Sigma_{xy} } ) \leq p$. Let  ${ \bf A } \in \mathbb{R}^{p \times r}$ and ${ \bf B } \in \mathbb{R}^{q \times r}$ be the matrices with in their columns the \textit{canonical vectors}. The new variables $\bf u ={\bf A}^T \bf x$ and $\bf v ={\bf B}^T \bf y$ are the \textit{canonical variates} and the correlations between each pair of canonical variates give the \textit{canonical correlations}. The canonical vectors contained in the matrices $\bf A$ and $\bf B$ are respectively given by the eigenvectors of the matrices
\begin{equation}
{\bf \Sigma_{xx}^{-1}}{\bf \Sigma_{xy}}{ \bf \Sigma_{yy}^{-1}}{ \bf \Sigma_{yx}} \text{\ \ \ \ and \ \ \ \ } { \bf \Sigma_{yy}^{-1} }{ \bf \Sigma_{yx}}{ \bf \Sigma_{xx}^{-1}}{ \bf \Sigma_{xy}}. \label{eigenvectors}
 \end{equation}
Both matrices have the same positive eigenvalues, the canonical correlations are given by the positive square root of those eigenvalues. 

The canonical vectors and correlations are typically estimated by taking the sample versions of the covariances in \eqref{eigenvectors} and computing the corresponding eigenvectors and eigenvalues. However, to implement this procedure, we need to invert the matrices $\widehat{ \bf \Sigma}_{\bf xx}$ and $\widehat{ \bf \Sigma}_{\bf yy}$. When the original variables are highly correlated or when the number of variables becomes large compared to the sample size, the estimation imprecision will be large. Moreover, when the largest number of variables in both data sets exceeds the sample size (i.e. $q \geq n$), traditional CCA cannot be performed. \cite{vinod76} proposed the canonical ridge, which is an adaptation of the ridge regression concept of \cite{Hoerl70} to the framework of CCA, to solve this problem. The canonical ridge replaces the matrices $\widehat{ \bf \Sigma}_{\bf xx}^{-1}$ and $\widehat{ \bf \Sigma}_{\bf yy}^{-1}$ by respectively ${(\widehat{ \bf \Sigma}_{\bf xx} + k_1 \bf I)}^{-1}$ and ${(\widehat{ \bf \Sigma}_{\bf yy} + k_2 \bf I )}^{-1}$. By adding the penalty terms $k_1$ and $k_2$ to the diagonal elements of the sample covariance matrices, one obtains more reliable and stable estimates when the data are nearly or exactly collinear.

Another approach is to use sparse CCA techniques. \cite{Parkhomenko09} consider a sparse singular value decomposition to derive sparse singular vectors. A limitation of their approach is that sparsity in the canonical vectors is only guaranteed if $\widehat{ \bf \Sigma}_{\bf xx}$ and $\widehat{ \bf \Sigma}_{\bf yy}$ are replaced by their corresponding diagonal matrices. A similar approach was taken by \cite{Witten09} who apply a penalized matrix decomposition to the cross-product matrix ${ \widehat{\bf \Sigma}_{xy} }$, but assume that one can replace the matrices $\widehat{ \bf \Sigma}_{\bf xx}$ and $\widehat{ \bf \Sigma}_{\bf yy}$ by identity matrices. \cite{Waaijenborg08} consider Wold's (\citeyear{wold1968}) alternating least squares approach to CCA and obtain sparse canonical vectors using penalized regression with elastic net. The ridge parameter of the elastic net is set to be large, thereby, according to the authors, ignoring the dependency structure within each set of variables.

\cite{Waaijenborg08}, \cite{Witten09},  and \cite{Parkhomenko09} all impose covariance restrictions, i.e. $\bf \Sigma_{XX} = \bf \Sigma_{YY} = I$ for \cite{Waaijenborg08} and \cite{Witten09}; diagonal matrices for \cite{Parkhomenko09}. In contrast, we propose in this paper to estimate the canonical variates without imposing any prior covariance restrictions. Our proposed method obtains the canonical vectors using a alternating penalized regression framework. By performing variable selection in a penalized regression framework using the lasso penalty \citep{Tibshirani96}, we obtain sparse canonical vectors.

We demonstrate in a simulation study that our Sparse Alternating Regression (SAR) algorithm produces good results in terms of estimation accuracy of the canonical vectors, and detection of the sparseness structure of the canonical vectors. Especially in simulation settings when there is a dependency structure within each set of variables, the SAR algorithm clearly outperforms the sparse CCA methods described above. We also apply the SAR algorithm on a high-dimensional genomic data set. Sparse estimation is appealing since it highlights the most important variables for the association study. 

The remainder of this article is organized as follows.  In Section 2 we formulate the CCA problem from a predictive point of view. Section 3 describes the Sparse  Alternating Regression (SAR) approach and provides details on the implementation of the algorithm. Section 4 compares our methodology to other sparse CCA techniques by means of a simulation study. Section 5 discusses the genomic data example, Section 6 concludes.

\section{CCA from a predictive point of view}
A characterization of the canonical vectors based on the concept of prediction is proposed by \cite{Brillinger75} and \cite{Izenman75}. Given $n$ observations ${\bf x_i} \in \mathbb{R}^{p}$ and  ${\bf y_i} \in \mathbb{R}^{q}$ ($i=1,\ldots,n$), consider the optimization problem 
\begin{equation}
(\widehat{{\bf A}}, \widehat{{\bf B}}) = \underset{({\bf A}, {\bf B}) \in \mathcal{S}}{\operatorname{argmin}} \ \sum_{i=1}^{n} || {\bf A}^T{\bf x_i} - {\bf B}^T {\bf y_i} ||^2. \label{objfunction}
\end{equation}
We restrict the parameter space to the space $\mathcal{S}$, given by
\small
\begin{equation}
\mathcal{S} = \{ {(\bf A, \bf B)} :{ \bf A } \in \mathbb{R}^{p \times r}, { \bf B } \in \mathbb{R}^{q \times r}, \text{rank}({\bf A})=\text{rank}({\bf B})=r , { \bf A^T } { \bf \Sigma_{xx} } { \bf A } = { \bf B^T } { \bf \Sigma_{yy} } { \bf B }= {\bf I}_r \}. \nonumber %\label{spaceS} 
\end{equation}
\normalsize
We impose normalization conditions requiring the canonical variates to have unit variance and to be uncorrelated.  \cite{Brillinger75} proves that the objective function in \eqref{objfunction} is minimized when $\bf A$ and $\bf B$ contain in their columns the canonical vectors. 

We build on this equivalent formulation of the CCA problem to obtain the canonical vectors using an alternating regression procedure (see e.g.  \citealp{wold1968}; \citealp{Branco05}). The subsequent canonical variates are sequentially derived. 

\bigskip

{\it First canonical vector pair.} Denote the first canonical vectors (i.e. the first columns of the matrices ${ \bf A }$ and ${ \bf B }$) by $({ \bf A_1 }, { \bf B_1 } )$.
Suppose we have an initial value $\bf A^*_1$ for the first canonical vector in the matrix $\bf A$. Then the minimization problem in \eqref{objfunction} reduces to 
\begin{equation}
\widehat{ \bf B}_1|{ \bf A^*_1}  = \underset{{\bf B_1}}{\operatorname{argmin}} \sum_{i=1}^{n} \left( {\bf A^*_1}^T{\bf x_i} - {\bf B_1}^T {\bf y_i} \right )^2, \label{objfunction_B1}
\end{equation}
where we require ${\bf \hat{v}_1}={\bf Y}\widehat{ \bf B}_1$ to have unit variance. The solution to \eqref{objfunction_B1} can be obtained from a multiple regression with ${\bf X}{\bf A^*_1}$ as response and ${\bf Y}$ as predictor, where ${\bf X}=[{\bf x_1},\ldots,{\bf x_n}]^{T}$ and ${\bf Y}=[{\bf y_1},\ldots,{\bf y_n}]^{T}$.

Analogously, for a fixed value $\bf B^*_1$. The optimal value for $\bf A_1$ is obtained by a multiple regression with ${\bf Y}{\bf B^*_1}$ as response and ${\bf X}$ as predictor
\begin{equation}
\widehat{ \bf A}_1|{ \bf B^*_1}  = \underset{{\bf A_1} }{\operatorname{argmin}} \sum_{i=1}^{n}\left({\bf B^*_1}^T{\bf y_i} - {\bf A_1}^T {\bf x_i} \right )^2, \label{objfunction_A1}
\end{equation}
where we require ${\bf \hat{u}_1}={\bf X}\widehat{ \bf A}_1$ to have unit variance. This leads to an alternating regression scheme, where we alternately update our estimates of the first canonical vectors until convergence. We iterate until the angle between the estimated canonical vectors in iteration $i$ and the respective estimated canonical vectors in the previous iteration are both smaller than some value $\epsilon$ (e.g. $\epsilon=10^{-3}$). 

\bigskip

{\it Higher order canonical vector pairs.} The higher order canonical variates need to be orthogonal to the previously found canonical variates. Therefore, the alternating regression scheme is applied on deflated data matrices (see e.g. \citealp{Branco05}). For the second pair of canonical vectors, consider the deflated matrices 
\begin{equation}
{\bf X}^{*} = {\bf X} - {\bf \hat{u}}_{1} ({\bf \hat{u}}^{T}_{1}{\bf \hat{u}}_{1})^{-1}{\bf \hat{u}}^{T}_{1}{\bf X}. \label{deflatedX}
\end{equation}
The deflated matrix ${\bf X}^{*}$ is obtained as the residuals of the multivariate regression of ${\bf X}$ on ${\bf \hat{u}}_{1}$, the first canonical variate.  Analogously, the deflated matrix ${\bf Y}^{*}$ is given by
\begin{equation}
{\bf Y}^{*} = {\bf Y} - {\bf \hat{v}}_{1} ({\bf \hat{v}}^{T}_{1}{\bf \hat{v}}_{1})^{-1}{\bf \hat{v}}^{T}_{1}{\bf Y}, \label{deflatedY}
\end{equation}
the residuals of the multivariate regression of ${\bf Y}$ on ${\bf \hat{v}}_{1}$. 

Using the Least Squares property, each column of ${\bf X}^{*}$ is uncorrelated with the first canonical variate ${\bf \hat{u}}_{1}$. The second canonical variate will be a linear combination of the columns of ${\bf X}^{*}$ and, hence, will be uncorrelated to the previously found canonical variate. The same holds for ${\bf Y}^{*}$.  The second canonical variate pair is then obtained by alternating between the following regressions until convergence:
\begin{equation}
\widehat{ \bf B}^*_2|{ \bf A^*_2}  = \underset{{\bf B^*_2}   }{\operatorname{argmin}} \sum_{i=1}^{n}\left( {\bf A^*_2}^T{\bf x^*_i} - {\bf B^*_2}^T {\bf y^*_i} \right)^2 \label{B2}
\end{equation} 
\begin{equation}
\widehat{ \bf A}^*_2|{ \bf B^*_2}  = \underset{{\bf A^*_2}  }{\operatorname{argmin}} \sum_{i=1}^{n}\left({\bf B^*_2}^T{\bf y^*_i} - {\bf A^*_2}^T {\bf x^*_i} \right)^2, \label{A2}
\end{equation}
where we require ${\bf \hat{v}^*_2}={\bf Y^*}\widehat{ \bf B}^*_2$  and ${\bf \hat{u}^*_2}={\bf X^*}\widehat{ \bf A}^*_2$ to have both unit variance.

Finally, we need to express the second canonical vector pair in terms of the original data sets ${\bf X}$ and ${\bf Y}$. To obtain the second canonical vector $\widehat{ \bf A}_2$, we regress ${\bf \hat{u}}^*_{2}$ on ${\bf X}$ 
\begin{equation}
\widehat{ \bf A}_2 = \underset{{\bf A_2}  }{\operatorname{argmin}} \sum_{i=1}^{n}\left({\bf \hat{u}}^*_{2} - {\bf A_2}^T {\bf x_i} \right)^2, \label{A2_final}
\end{equation}
yielding the fitted values ${\bf \hat{u}}_{2}={\bf X}\widehat{ \bf A}_2$. To obtain $\widehat{ \bf B}_2$, we regress ${\bf \hat{v}}^*_{2}$ on ${\bf Y}$.  
\begin{equation}
\widehat{ \bf B}_2 = \underset{{\bf B_2}  }{\operatorname{argmin}} \sum_{i=1}^{n}\left({\bf \hat{v}}^*_{2} - {\bf B_2}^T {\bf y_i} \right)^2. \label{B2_final}
\end{equation}
The same idea is applied to obtain the higher order canonical variate pairs. 

\section{Sparse alternating regressions}
The canonical vectors obtained with the alternating regression scheme from Section 2 are in general not sparse. Sparse canonical vectors are obtained by replacing the Least Squares regressions in the alternating regression approach of Section 2 with Lasso regressions ($L_1$-penalty). As such, some coefficients in the canonical vectors will be set to exactly zero, thereby producing linear combinations of only a subset of variables.

For the first pair of sparse canonical vectors, the sparse equivalents of the Least Squares regressions in equations \eqref{objfunction_B1} and \eqref{objfunction_A1} are given by
\begin{equation}
\widehat{ \bf B}_1|{ \bf A^*_1}  = \underset{{\bf B_1}}{\operatorname{argmin}} \sum_{i=1}^{n} \left( {\bf A^*_1}^T{\bf x_i} - {\bf B_1}^T {\bf y_i} \right )^2 + \lambda_{{\bf B}_1}\sum_{j=1}^{q}|{\bf b}_{j1}|, \nonumber %\label{sparse_B1}
\end{equation}
\begin{equation}
\widehat{ \bf A}_1|{ \bf B^*_1}  = \underset{{\bf A_1} }{\operatorname{argmin}} \sum_{i=1}^{n}\left({\bf B^*_1}^T{\bf y_i} - {\bf a_1}^T {\bf x_i} \right )^2 + \lambda_{{\bf A}_1}\sum_{j=1}^{p}|{\bf a}_{j1}|, \nonumber %\label{sparse_A1}
\end{equation}
where $\lambda_{{\bf B}_1}>0$ and $\lambda_{{\bf A}_1}>0$ are sparsity parameters, ${\bf b}_{j1}$ is the $j^{th}$ ($j=1,\ldots,q$) element of the first canonical vector ${\bf B}_{1}$ and ${\bf a}_{j1}$ is the $j^{th}$ ($j=1,\ldots,p$) element of the first canonical vector ${\bf A}_{1}$. The first pair of canonical variates are given by  ${\bf \hat{u}_1}={\bf X}\widehat{ \bf A}_1$ and ${\bf \hat{v}_1}={\bf Y}\widehat{ \bf B}_1$. We require both to have unit variance.

To obtain the second pair of sparse canonical vectors, the same deflated matrices as in equations \eqref{deflatedX} and \eqref{deflatedY} are used.  The Least Squares regressions in equations \eqref{B2} and \eqref{A2} are replaced by the Lasso regressions
\begin{equation}
\widehat{ \bf B}^*_2|{ \bf A^*_2}  = \underset{{\bf B^*_2}   }{\operatorname{argmin}} \sum_{i=1}^{n}\left( {\bf A^*_2}^T{\bf x^*_i} - {\bf B^*_2}^T {\bf y^*_i} \right)^2 + \lambda_{{\bf B}^*_2}\sum_{j=1}^{q}|{\bf b}^{*}_{j2}| \nonumber %\label{sparseB2}
\end{equation} 
\begin{equation}
\widehat{ \bf A}^*_2|{ \bf B^*_2}  = \underset{{\bf A^*_2}  }{\operatorname{argmin}} \sum_{i=1}^{n}\left({\bf B^*_2}^T{\bf y^*_i} - {\bf A^*_2}^T {\bf x^*_i} \right)^2 + \lambda_{{\bf A}^*_2}\sum_{j=1}^{p}|{\bf a}^{*}_{j2}|. \nonumber %\label{sparseB1}
\end{equation}
Finally, to express the second pair of canonical vectors in terms of the original data matrices, we replace the Least Squares regression in \eqref{A2_final} and \eqref{B2_final} by the two Lasso regressions. 
\begin{equation}
\widehat{ \bf A}_2 = \underset{{\bf A_2}  }{\operatorname{argmin}} \sum_{i=1}^{n}\left({\bf \hat{u}}^*_{2} - {\bf A_2}^T {\bf x_i} \right)^2 + \lambda_{{\bf A}_2}\sum_{j=1}^{p}|{\bf a}_{j2}| , \nonumber %\label{A2_finalsparse}
\end{equation}
\begin{equation}
\widehat{ \bf B}_2 = \underset{{\bf B_2}  }{\operatorname{argmin}} \sum_{i=1}^{n}\left({\bf \hat{v}}^*_{2} - {\bf B_2}^T {\bf y_i} \right)^2 + \lambda_{{\bf B}_2}\sum_{j=1}^{q}|{\bf b}_{j2}|, \nonumber %\label{B2_finalsparse}
\end{equation}
yielding the fitted values ${\bf \hat{u}}_{2}={\bf X}\widehat{ \bf A}_2$ and ${\bf \hat{v}}_{2}={\bf Y}\widehat{ \bf B}_2$. We add a lasso penalty to the above regressions, first because the design matrix $\bf{X}$ can be high-dimensional, and second, because we want $\widehat{ \bf A}_2$ and $\widehat{ \bf B}_2$ to be sparse.

A complete description of the algorithm is given below. We numerically verified that without imposing penalization (i.e. $\lambda_{A,j}=\lambda_{B,j}=0$, for $j=1,\ldots,r$), the traditional CCA solution is obtained. Our numerical experiments all converged reasonably fast. Finally, note that as in other sparse CCA proposals (\citealp{Witten09}; \citealp{Parkhomenko09}; \citealp{Waaijenborg08}) the sparse canonical variates are in general not uncorrelated. We do not consider this lack of uncorrelatedness as a major flaw. The sparse canonical vectors yield an easily interpretable basis of the space spanned by the canonical vectors. After suitable rotation of the corresponding canonical variates, this basis can be made orthogonal (but not sparse) if one  desires so.

\spacingset{1}
\footnotesize
\begin{center} 
\fbox{\begin{minipage}{6.5in} 
\textbf{Sparse Alternating Regression (SAR) Algorithm } 
\bigskip

Let ${\bf X}$ and ${\bf Y}$ be two data matrices.
\begin{enumerate}
\item \textit{Preliminary steps}
	\begin{itemize}
	\item ${\bf X}_0 = {\bf X} - {\bf 1}{\bf \bar{x}}^{T}$
	\item ${\bf Y}_0 = {\bf Y} - {\bf 1}{\bf \bar{y}}^{T} $
	\end{itemize}
\item \textit{Alternating Regressions}: For $l=1,\ldots,r$
\begin{itemize}
	\item If $l > 1$ : \textit{Deflated matrices}
		\begin{itemize} 
		\item[$\bullet$] ${\bf X}_{l-1} = {\bf X}_{l-2} - {\bf \hat{u}}_{l-1} ({\bf \hat{u}}^{T}_{l-1}{\bf \hat{u}}_{l-1})^{-1}{\bf \hat{u}}^{T}_{l-1}{\bf X}_{l-2}$
		\item[$\bullet$] ${\bf Y}_{l-1} = {\bf Y}_{l-2} - {\bf \hat{v}}_{l-1} ({\bf \hat{v}}^{T}_{l-1}{\bf \hat{v}}_{l-1})^{-1}{\bf \hat{v}}^{T}_{l-1}{\bf Y}_{l-2}$
		\end{itemize}
	\item  \textit{Starting values}  
		\begin{itemize} 
		\item[$\bullet$] ${\bf {\widehat{B}}}_l^{(0)} = \frac{\hat{b}^{can ridge}_l}{|| \hat{b}^{can ridge}_l || } $, using the canonical vector $\hat{b}^{can ridge}_l$ obtained with the canonical ridge 
		\item[$\bullet$] ${\bf {\widehat{v}}}_l^{(0)} = {\bf Y}_{l-1}{\bf {\widehat{B}}}_l^{(0)} $ 
		\end{itemize}
\item \textit{From iteration $s=1$ until convergence}
		\begin{itemize} 
		\item[$\bullet$] ${\bf {\widehat{a}}}_l^{(s)}=\underset{a}{\operatorname{argmin}} \sum_{i=1}^{n} \bigl ({\bf {\widehat{v}}}_{l,i}^{(s-1)}  - {\bf x}_{l-1, i}^{T}a \bigr )^2  + \lambda_{a,l} \sum_{j=1}^{p} |a_j|$  
		\item[$\bullet$] ${\bf {\widehat{a}}}_l^{(s)} = \frac{{\bf {\widehat{a}}}_l^{(s)}}{|| {\bf {\widehat{a}}}_l^{(s)} || } $
		\item[$\bullet$] ${\bf {\widehat{u}}}_l^{(s)} = {\bf X}_{l-1}{\bf {\widehat{a}}}_l^{(s)} $ 
		\item[$\bullet$] ${\bf {\widehat{b}}}_l^{(s)}=\underset{b}{\operatorname{argmin}} \sum_{i=1}^{n} \bigl ({\bf {\widehat{u}}}_{l,i}^{(s)}  - {\bf y}_{l-1, i}^{T}b \bigr )^2  + \lambda_{b,l} \sum_{j=1}^{q} |b_j|$  
		\item[$\bullet$] ${\bf {\widehat{b}}}_l^{(s)} = \frac{{\bf {\widehat{b}}}_l^{(s)}}{|| {\bf {\widehat{b}}}_l^{(s)} || } $
		\item[$\bullet$] ${\bf {\widehat{v}}}_l^{(s)} = {\bf Y}_{l-1}{\bf {\widehat{b}}}_l^{(s)} $  
		\end{itemize}
\item \textit{After convergence}, resulting in ${\bf {\widehat{a}}}^{*}_l, {\bf {\widehat{b}}}^{*}_l, {\bf {\widehat{u}}}^{*}_l$ and ${\bf {\widehat{v}}}^{*}_l$
		\begin{itemize} 
	     \item[$\bullet$] ${\bf \widehat{U}}_{l-1}= [\widehat{u}_1,\ldots,\widehat{u}_{l-1}]$
	     \item[$\bullet$] $\tilde{u}_l = {\bf {\widehat{u}}}^{*}_l - {\bf \widehat{U}}_{l-1} \left( {\bf \widehat{U}}_{l-1}^T{\bf \widehat{U}}_{l-1}  \right)^{-1}{\bf \widehat{U}}_{l-1}^T{\bf {\widehat{u}}}^{*}_l $
	     \item[$\bullet$] ${\bf {\widehat{A}}}_l = \begin{cases}
		 {\bf {\widehat{a}}}^{*}_l & \textbf{ if  } l=1 \\
		 \underset{A}{\operatorname{argmin}} \sum_{i=1}^{n} \bigl ({\bf {\tilde{u}}}_{l,i}  - {\bf x}_{0, i}^{T}A \bigr )^2  + \lambda_{A,l} \sum_{j=1}^{p} |A_j| & \textbf{ if  } l>1
		 \end{cases}$
		 \item[$\bullet$] $\widehat{u}_l={\bf X}_0{\bf {\widehat{A}}}_l$
		 \item[$\bullet$] ${\bf \widehat{V}}_{l-1}= [\widehat{v}_1,\ldots,\widehat{v}_{l-1}]$
		 \item[$\bullet$] $\tilde{v}_l = {\bf {\widehat{v}}}^{*}_l - {\bf \widehat{V}}_{l-1} \left( {\bf \widehat{V}}_{l-1}^T{\bf \widehat{V}}_{l-1}  \right)^{-1}{\bf \widehat{V}}_{l-1}^T{\bf {\widehat{v}}}^{*}_l $
		 \item[$\bullet$] ${\bf {\widehat{B}}}_l = \begin{cases}
		 {\bf {\widehat{b}}}^{*} & \textbf{ if  } l=1 \\
		 \underset{B}{\operatorname{argmin}} \sum_{i=1}^{n} \bigl ({\bf {\tilde{v}}}_{l,i}  - {\bf y}_{0, i}^{T}B \bigr )^2  + \lambda_{B,l} \sum_{j=1}^{q} |B_j|& \textbf{ if  } l>1
		 \end{cases}$
		 \item[$\bullet$] $\widehat{v}_l={\bf Y}_0{\bf {\widehat{B}}}_l$
		\end{itemize}
\end{itemize}
\item \textit{Final solution}
		\begin{itemize} 
		\item[$\bullet$] $\widehat{\bf A}_{\text{sparse}} =  [{\bf {\widehat{A}}}_1,\ldots,{\bf {\widehat{A}}}_r] $
		\item[$\bullet$] $\widehat{\bf B}_{\text{sparse}} =  [{\bf {\widehat{B}}}_1,\ldots,{\bf {\widehat{B}}}_r] $
		\end{itemize}
\end{enumerate}
\end{minipage}}
\end{center}
\spacingset{1.45}
\normalsize
\bigskip

\bigskip

{ \it Starting values.} To start up the Sparse Alternating Regression (SAR) algorithm, an initial value is required. We use the  canonical vectors delivered by the canonical ridge as starting value, which is available at no computational cost. The regularization parameters of the canonical ridge are chosen using $5$-fold cross-validation such that the average test sample canonical correlation is maximized \citep{Gonzalez08}. 

We performed a simulation study (unreported) to assess the robustness of the SAR algorithm to different choices of starting values. The SAR algorithm shows similar performance when either the canonical ridge or other choices of starting values (i.e. CCA in low-dimensional settings and randomly drawn starting values) are used. 

\bigskip

{ \it Number of canonical variates to extract.} For practical implementation, one needs to have an idea on the number of canonical variates $r$ to extract. Most often, only a limited number of canonical variate pairs are truly relevant. We follow \cite{An13} who propose the maximum eigenvalue ratio criterion to decide on the number of canonical variates to extract. We apply the canonical ridge and calculate the canonical correlations $\hat{\rho}_1, \ldots,\hat{\rho}_{rmax} $, with $rmax=min(p,q)$. Let $\hat{k}_j = \hat{\rho}_j/\hat{\rho}_{j+1}$ for $j=1,\ldots,rmax-1$.  Then we set $r=\text{argmax}_j \hat{k}_j$, and extract $r$ pairs of canonical variates using the SAR algorithm.

\bigskip

{\it Selection of sparsity parameters.} In the SAR algorithm,  the sparsity parameters $\lambda_{A,j}$ and $\lambda_{B,j}$ $(j=1,\ldots,r)$, which control the penalization on the respective regression coefficient matrices, need to be selected.  We select the sparsity parameters according to a minimal Bayes Information Criterion (BIC).
We solve the corresponding penalized regression problems over a range of values and select for each the one with lowest value of
\begin{equation}\label{eq: BICbetaA}
BIC_{\lambda_{A,j}} = -2 \log L_{\lambda_{A,j}} + k_{\lambda_{A,j}} \log(n), \nonumber
\end{equation}
\begin{equation}\label{eq: BICbetaB}
BIC_{\lambda_{B,j}} = -2 \log L_{\lambda_{B,j}} + k_{\lambda_{B,j}} \log(n), \nonumber
\end{equation}
for $j=1,\ldots,r$. $L_{\lambda_{A,j}}$ is the estimated likelihood using sparsity parameter $\lambda_{A,j}$ and $k_{\lambda_{A,j}}$ is the number of non-zero estimated regression coefficients. Analogously for $\lambda_{B,j}$.

\section{Simulation Study}
We compare the performance of the Sparse Alternating Regression approach with three other sparse CCA techniques. We consider
\begin{itemize}
\item The Sparse Alternating Regression (SAR) algorithm detailed in Section 3.
\item The sparse CCA of \cite{Witten09}\footnote{Available in the R package \texttt{PMA} \citep{Rpma}.}, relying on a penalized matrix decomposition applied to the cross-product matrix ${ \widehat{\bf \Sigma}_{xy} }$. Sparsity parameters are selected using the permutation approach described in \cite{Correlate}.
\item The sparse CCA of \cite{Parkhomenko09}\footnote{Available at http://www.uhnres.utoronto.ca/labs/tritchler/.}. Sparsity parameters are selected using $5$-fold cross-validation where the average test sample canonical correlation is maximized.
\item The sparse CCA of \cite{Waaijenborg08}\footnote{We re-implemented the algorithm of \cite{Waaijenborg08} in R.}.  The lasso parameter of the elastic net is selected using $5$-fold cross-validation such that the mean absolute difference between the canonical correlation of the training and test sets is minimized.
\end{itemize}
We emphasize that the sparsity parameters of all methods are selected as proposed by the respective authors. The traditional CCA solution and the canonical ridge\footnote{Available in the R package \texttt{CCA} \citep{Rcca}.} are computed as additional benchmarks. 

We consider several simulation schemes. For each setting we generate data matrices $\bf X$ and $\bf Y$ according to multivariate normal distributions, with covariance matrices described in Table \ref{Sim_setup}. The number of simulations for each setting is $M=1000$. In all simulation settings, the canonical vectors have a sparse structure. In the first simulation setup (revised from \citealp{Branco05}) the covariance restrictions of \cite{Waaijenborg08}, \cite{Witten09} and \cite{Parkhomenko09} (i.e. $\bf \Sigma_{XX} = \bf \Sigma_{YY} = I$ for the former two, diagonal matrices for the latter) are satisfied. These restrictions are violated in the second, third and fourth simulation setup. In the third design, the number of variables is large compared to the sample size. Traditional CCA can still be performed in this setting. In the fourth design, the number of variables in the data matrix  $\bf Y$ is larger than the sample size, and traditional CCA can no longer be performed.

\begin{table}
\footnotesize
\caption{Simulation settings}\label{Sim_setup}
\begin{center}
\begin{tabular} {l|cccccc}
\hline
Design & $n$ &$p$ & $q$ & $ { \bf \Sigma_{xx} }$ & $ { \bf \Sigma_{yy} }$ & $ { \bf \Sigma_{xy} }$  \\ \hline
&&&&&&\\
Uncorrelated  & 50 & 4 & 6 & $\bf {I}_p$ & $ \bf  {I}_q$ & $\begin{bmatrix}  \dfrac{3}{5} & 0 & 0 & 0 & 0 & 0 \\ 0 & \dfrac{1}{2} & 0 & 0 & 0 & 0 \\ 0 & 0 & 0 & 0 & 0 & 0 \\ 0 & 0 & 0 & 0 & 0 & 0 \end{bmatrix}$ \\
&&&&&&\\
&&&&&&\\
Correlated & 50 & 6 & 10 & $\bf {I}_p$ &  $ \begin{bmatrix}  \bf{S_1} & \bf{0} \\ \bf{0} & \bf {I}_7 \end{bmatrix}$ &  $\begin{bmatrix}  \bf{S_2} & \bf{0} \\ \bf{0} & \bf{0} \end{bmatrix}$ \\
&&&& \text{with} &  & \\
&&&&& ${{\bf{S_1}}_{ij}}= 0.7^{\mid i-j \mid} $& ${\bf{S_2}}= \frac{1}{2} \bf {I}_2 $\\
&&&&&&\\
&&&&&&\\
High-dimensional & 50 & 25 & 40 & $\bf {I}_p$ &  $ \begin{bmatrix}  \bf{S_1} & \bf{0} \\ \bf{0} & \bf {I}_{37} \end{bmatrix}$ &  $\begin{bmatrix}  \bf{S_2} & \bf{0} \\ \bf{0} & \bf{0} \end{bmatrix}$ \\
&&&& \text{with} &  & \\
&&&&& ${{\bf{S_1}}_{ij}}= 0.3^{\mid i-j \mid} $& ${\bf{S_2}}= \frac{7}{10} \bf {I}_2 $\\
&&&&&&\\
&&&&&&\\
Overparametrized & 80 & 60 & 85 & $\bf {I}_p$ &  $ \begin{bmatrix}  \bf{S_1} & \bf{0} \\ \bf{0} & \bf {I}_{82} \end{bmatrix}$ &  $\begin{bmatrix}  \bf{S_2} & \bf{0} \\ \bf{0} & \bf{0} \end{bmatrix}$ \\
&&&& \text{with} &  & \\
&&&&& ${{\bf{S_1}}_{ij}}= 0.3^{\mid i-j \mid} $& ${\bf{S_2}}= \frac{7}{10} \bf {I}_2 $\\

&&&&&&\\ \hline
\end{tabular}
\end{center}
\end{table}

\bigskip

{\it Performance measures.} We compare the SAR algorithm to its alternatives and evaluate (i) the precision accuracy of the space spanned by the estimated canonical vectors, and (ii) the detection of the sparsity structure of the canonical vectors.

We compute for each simulation run $m$, with $m=1,\ldots,M$, the angle $\theta^m({\bf \hat{A}}^m,{\bf A})$ between the subspace spanned by the estimated canonical vectors contained in the columns of ${\bf \hat{A}}^m$ and the subspace spanned by the true canonical vectors contained in the columns of $\bf A$. Analogously for the matrix $\bf B$. The average angles are given by
\begin{equation}
\theta({\bf \hat{A}},{\bf A}) = \dfrac{1}{M} \sum_{m=1}^{M} \theta^m({\bf \hat{A}}^m,{\bf A}) \text{\ \ \ and \ \ \ \ }
\theta({\bf \hat{B}},{\bf B}) = \dfrac{1}{M} \sum_{m=1}^{M} \theta^m({\bf \hat{B}}^m,{\bf B}). \nonumber
\end{equation}

Finally, we monitor the sparsity recognition performance (e.g. \citealp{Rothman10}) using the true positive rate and the true negative rate as defined as follows
\begin{gather}
TPR({\bf \hat{A}},{\bf A}) = \frac{ \# \{ (i,j):{\bf {\widehat{A_{ij}}}} \neq 0 \text{\ \ and \ }  {\bf {A_{ij}}} \neq 0   \}} { \# \{ (i,j):  {\bf {A_{ij}}} \neq 0         \}} \nonumber \\
TNR({\bf \hat{A}},{\bf A}) = \frac{ \# \{ (i,j):{\bf {\widehat{A_{ij}}}} = 0 \text{\ \ and \ }  {\bf {A_{ij}}} = 0   \}} { \# \{ (i,j):  {\bf {A_{ij}}} = 0         \}}. \nonumber 
\label{sparsityperformance} 
\end{gather}
The true positive rate indicates the number of true relevant variables detected by the estimation procedure. The true negative rate measures the hit rate of excluding unimportant variables from the canonical vectors.  Analogue measures can be computed for the canonical vectors in the matrix $\bf B$.

\bigskip

{\it Results.} The simulation results on the estimation accuracy of the estimated canonical vectors are reported in Table \ref{Sim1_estimaccuracy}. We compute the average angle (averaged across simulation runs) between the space spanned by the true and estimated canonical vectors. To compare the average angle of the SAR algorithm against the other approaches, we compute $p$-values of a two-sided paired $t$-test.

\begin{table}
\small
\caption{Estimation accuracy of the canonical vectors, measured by the average angle between the subspace spanned by the true and estimated canonical vectors. $P$-values comparing SAR to alternatives are all $<0.01$. }\label{Sim1_estimaccuracy}
\begin{center}
\begin{tabular} {llllllccccc}
\hline
Design &&& Method &&&& $\theta({\bf \hat{A}},{\bf A})$ &&& $\theta({\bf \hat{B}},{\bf B})$    \\ \hline
Uncorrelated&&& SAR &&&& 0.008&&&  0.019 \\
   &&& \cite{Witten09}&&&& 0.010&&&  0.054\\ 
    &&& \cite{Parkhomenko09} &&&& 0.104 &&& 0.241  \\
  &&& \cite{Waaijenborg08}&&&& 0.078 &&& 0.212 \\
   &&& Canonical ridge  &&&& 0.129 &&&  0.272  \\
   &&& CCA &&&& 0.127 &&&  0.267  \\
 &&&&  &&&&&  &  \\
Correlated &&& SAR &&&& 0.001 &&& 0.068  \\
     &&& \cite{Witten09} &&&& 0.061 &&&  0.299  \\ 
       &&& \cite{Parkhomenko09} &&&& 0.306 &&& 0.690  \\
   &&& \cite{Waaijenborg08}&&&& 0.187 &&& 0.494  \\
     &&& Canonical ridge  &&&& 0.046 &&& 0.043  \\
     &&& CCA &&&& 0.042 &&&  0.033  \\
  &&&&  &   &&&&&  \\
High-dimensional &&& SAR &&&& 0.212&&&  0.305 \\
   &&& \cite{Witten09}&&&& 0.263 &&&  0.390 \\
    &&& \cite{Parkhomenko09} &&&& 0.826 &&& 0.908  \\
  &&& \cite{Waaijenborg08} &&&& 0.833 &&& 0.942  \\ 
   &&& Canonical ridge  &&&& 0.916 &&&  1.016   \\
   &&& CCA &&&& 1.062 &&& 1.193  \\ 
   
     &&&&  &   &&&&&  \\

Overparametrized &&& SAR &&&& 0.278 &&&  0.353  \\
      &&& \cite{Witten09}&&&& 0.490 &&&  0.592  \\
       &&& \cite{Parkhomenko09} &&&& 0.946 &&& 0.986 \\
     &&& \cite{Waaijenborg08} &&&& 1.137 &&& 1.181  \\ 
      &&& Canonical ridge  &&&& 0.916 &&& 1.211   \\ \hline
\end{tabular}
\end{center}
\end{table}

We first compare the performance of the penalized CCA techniques (i.e. canonical ridge and sparse CCA) to the unpenalized CCA solution. The estimation accuracy of the penalized CCA methods is significantly better compared to traditional CCA, especially in the high-dimensional design. In the lower dimensional simulation settings (i.e. uncorrelated and correlated design), sparse CCA techniques are still doing well since the underlying structure of the canonical vectors is sparse. %Overall, the SAR algorithm produces significantly more precise estimates of the canonical vectors compared to the traditional CCA solution. Furthermore, the sparse CCA techniques perform, in general, significantly better than the canonical ridge.

Next, we compare the SAR algorithm to its sparse alternatives. In the uncorrelated design, the covariance restrictions imposed by \cite{Waaijenborg08}, \cite{Parkhomenko09} and \cite{Witten09} are satisfied. Therefore, we expect these methods to perform especially well. Nevertheless, even in this setting, the SAR algorithm performs significantly better. In the correlated design, the high-dimensional and the overparametrized design these covariance restrictions are violated. Here, we see even more clearly that the SAR algorithm has a significant advantage over its sparse alternatives. In the correlated design, for instance, the SAR algorithm outperforms the method of \cite{Witten09} by a factor 10 for the first canonical vector (i.e. estimation accuracy of 0.001 against 0.061), and by a factor 5 for the second canonical vector (i.e. estimation accuracy of 0.068 against 0.299). The gains in estimation accuracy of the SAR algorithm compared to the other sparse CCA methods are even more outspoken.  %Hence, it appears to be less affected by the underlying structure of the covariance matrices compared to the other sparse CCA techniques that do impose quite restrictive covariance assumptions.

Finally, Table \ref{Sim1_SRP} compares the results on sparsity recognition performance among the sparse CCA techniques. The methods of \cite{Parkhomenko09} and \cite{Waaijenborg08} produce the least sparse solution, indicated by the high true positive rates and low true negative rates. The SAR algorithm and the method of \cite{Witten09} tend to produce the most spare solutions, indicated by the high true negative rates and low true positive rates. Contrary to sparse CCA, traditional CCA and the canonical ridge don't perform variable selection simultaneously with model estimation. Therefore, traditional CCA and canonical ridge are not included in Table \ref{Sim1_SRP}. All elements of the canonical vectors are estimated as non-zero, resulting in a perfect true positive rate and zero true negative rate. 

To conclude, as we can see from Table \ref{Sim1_estimaccuracy}, in every simulation design we consider, the SAR algorithm did perform significantly better than the other sparse CCA methods.

\begin{table}
\small
\caption{Sparsity recognition performance: true positive rate and true negative rate for canonical vectors in the $\bf A$ and $\bf B$ matrices. }\label{Sim1_SRP}
\begin{center}
\begin{tabular} {llccccccc}
\hline
& && \multicolumn{2}{c}{$\bf A$}  &&&  \multicolumn{2}{c}{$\bf B$}  \\ 
Design & Method && TPR & TNR &&&  TPR & TNR  \\ \hline
Uncorrelated & SAR && 0.79 & 0.81 &&& 0.79 & 0.86 \\
  & \cite{Witten09} && 0.75 & 0.83 &&& 0.77 & 0.78 \\
   & \cite{Parkhomenko09} && 0.94 & 0.22&&& 0.93 & 0.25\\
  & \cite{Waaijenborg08} &&  0.91 & 0.25 &&& 0.91 & 0.25\\ 
 &&  &&&   &&& \\

Correlated & SAR && 0.83 & 0.92 &&& 0.55 & 0.93\\
  & \cite{Witten09} && 0.52 & 0.80 &&& 0.43 & 0.75 \\ 
   & \cite{Parkhomenko09} &&  0.86 &  0.23 &&& 0.83 &  0.26\\
  & \cite{Waaijenborg08} && 0.86 & 0.31 &&& 0.83 & 0.31 \\ 
 &&  &&&&&&     \\
 
High-dimensional & SAR && 0.54 & 0.82 &&& 0.51 & 0.83\\
  & \cite{Witten09} && 0.38 & 0.87 &&& 0.31&  0.86\\ 
   & \cite{Parkhomenko09} && 0.72 & 0.35 &&& 0.70&  0.40 \\
  & \cite{Waaijenborg08} && 0.87 & 0.24 &&& 0.82 &0.27 \\ 
   &&  &&&&&&     \\

Overparametrized & SAR && 0.44 & 0.89 &&& 0.44 & 0.89 \\
    & \cite{Witten09} && 0.37  & 0.82 &&& 0.32 & 0.82 \\ 
     & \cite{Parkhomenko09} && 0.61 &  0.49 &&& 0.56 & 0.55 \\
    & \cite{Waaijenborg08} && 0.90  & 0.16  &&& 0.87 & 0.18\\\hline

\end{tabular}
\end{center}
\end{table}

\section{Genomic data application}
In recent years, high-dimensional genomic data sets have arisen, containing thousands of gene expression and other phenotype measurements (e.g., \citealp{Chen10, Daye12}). We use the publicly available breast cancer data set described in \cite{Chin06} and available in the R package \texttt{PMA} \citep{Rpma}. Comparative genomic hybridization (CGH) data (2149 variables) and gene expression data (19 672 variables) are available on 89 samples. The objective is to identify copy number change variables that are correlated with a subset of gene expression variables. Copy number changes on a particular chromosome are associated with expression changes in genes located on the same chromosome \citep{Witten09}. Therefore, we analyze the data for each chromosome separately, each time using the CGH and gene expression variables for that particular chromosome. The dimension of both sets of variables is large compared to the sample size such that traditional CCA cannot be performed. In such high-dimensional setting, the use of sparse CCA techniques is appealing.   We use the SAR algorithm to perform sparse CCA for each chromosome separately. 

\begin{figure}
\begin{center}
\includegraphics[width=15cm]{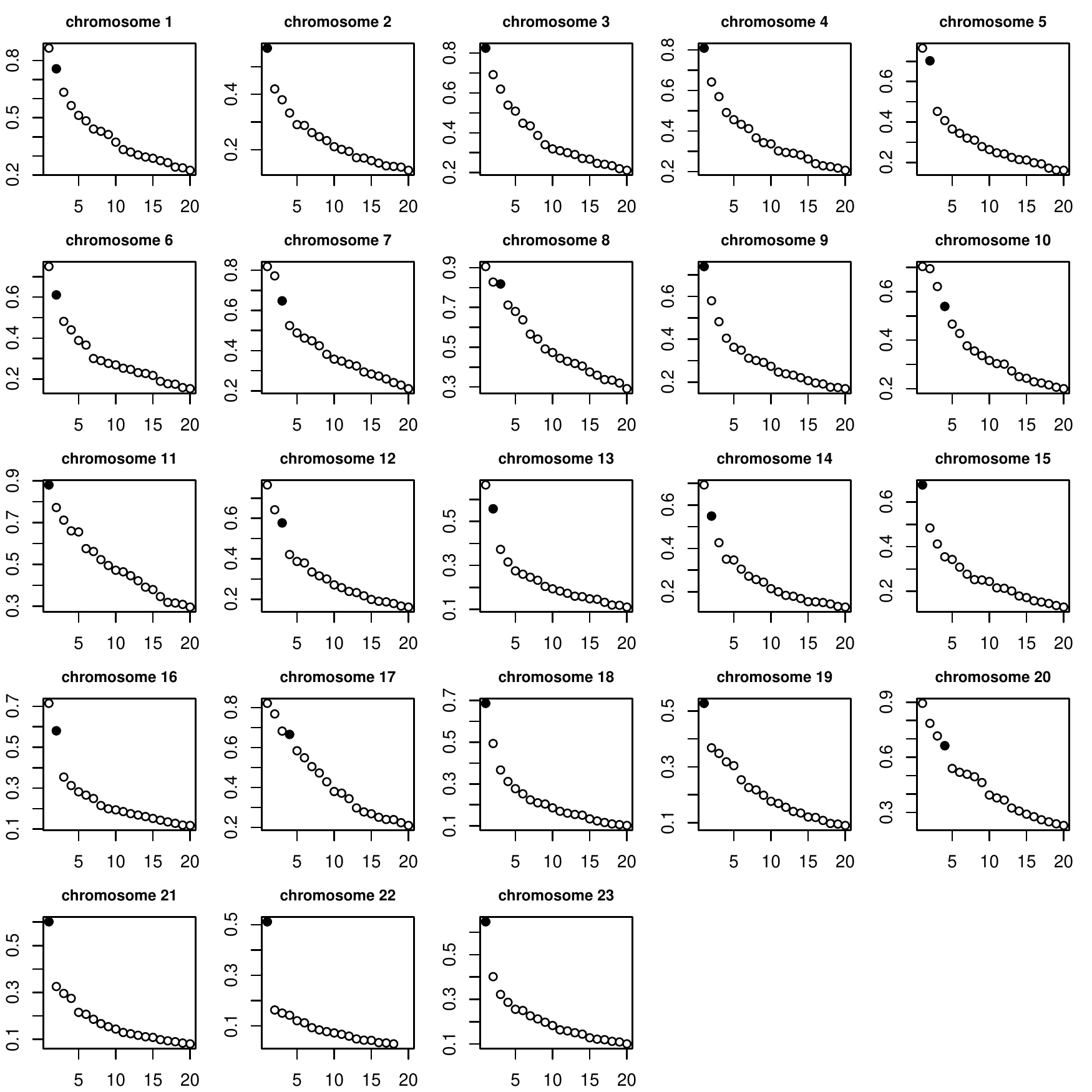} 
\caption{Estimated canonical correlations using the canonical ridge, for each of the 23 chromosomes. The highest order pair of  canonical variates to retain, as selected by the  maximum eigenvalue ratio criterion, is indicated by a solid black circle.}  \label{CR_cor}
\end{center}
\end{figure}

To decide on the number of canonical variates pairs to extract, we apply the canonical ridge to each chromosome. Figure \ref{CR_cor} shows the first 20 estimated canonical correlations for each of the 23 chromosomes. For each chromosome, we use the maximum eigenvalue ratio criterion, discussed in Section 3, to determine the number of canonical variate pairs to extract. Depending on the specific chromosome, this criterion indicates to extract either 1, 2, 3 or 4 canonical variate pairs. 

To compare the performance of the SAR algorithm to the other sparse CCA procedures discussed in Section 4, we perform an out-of-sample cross-validation exercise. More precisely, we perform a leave-one-out cross-validation exercise and compute the cross-validation score
\begin{equation}
CV = \frac{1}{n} \sum_{i=1}^{n} || {\bf \widehat{A}}^{T}_{-i}{\bf x}_i  - {\bf \widehat{B}}^{T}_{-i}{\bf y}_i ||^2, \nonumber
\end{equation}
where ${\bf \widehat{A}}^{T}_{-i}$ and ${\bf \widehat{B}}^{T}_{-i}$  contain the estimated canonical vectors when the $i^{th}$ observation is left out of the estimation sample. We compute this cross-validation score for each of the sparse CCA techniques. The technique that leads to the lowest value of this cross-validation score achieves the best out-of-sample performance.

\begin{figure}
\begin{center}
\includegraphics[width=12cm]{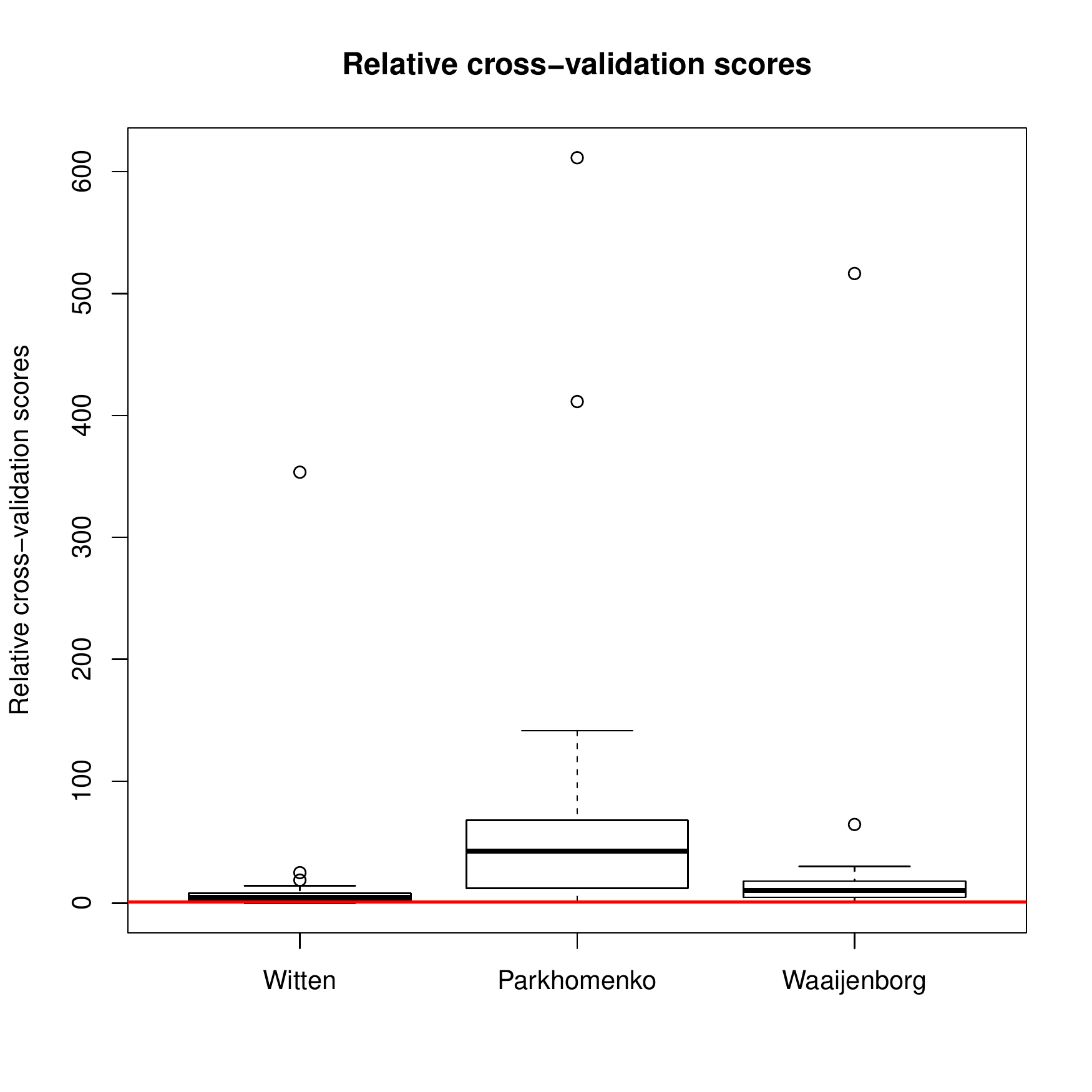} 
\caption{Boxplots of the 23 relative cross-validation scores  of \cite{Witten09}, \cite{Parkhomenko09} and \citeauthor{Waaijenborg08}, relative to the SAR algorithm. \label{CVapplication}}
\end{center}
\end{figure}

Averaged across all chromosomes, the SAR algorithm attains a cross-validation score of 87.21, the method of \cite{Witten09} 367.38, \cite{Parkhomenko09} 2778.57 and \cite{Waaijenborg08} 713.57. Thus, the SAR algorithm outperforms its alternatives. Furthermore, we compute relative cross-validation scores, being the cross-validation score of a method relative to the cross-validation score of the SAR algorithm. Boxplots of these relative cross-validations scores (23 scores, one for each chromosome) are presented in Figure  \ref{CVapplication}. A value of the relative cross-validation score larger than 1 (horizontal red line) indicates better performance of the SAR algorithm. The SAR algorithm always attains the best cross-validation score, except for two cases (out of 23) where \cite{Witten09} achieves the lowest cross-validation score. The differences in performance compared to the method of \cite{Parkhomenko09} and \cite{Waaijenborg08} are large. The cross-validation scores obtained with the SAR algorithm and the method of \cite{Witten09} are substantially lower than those obtained with the method of \cite{Parkhomenko09} and \cite{Waaijenborg08}. The solutions obtained with the former two are much sparser than the once obtained with the later two. Sparsity thus helps in achieving a good cross-validation score.

The dependency structure within each set of variables might explain the good performance of the SAR algorithm relative to its alternatives. For the first chromosome, for instance, 20\% of the (absolute) correlations between the 136 CGH spots are larger than 0.6. The same holds for the other chromosomes. In the simulation study from Section 4, we show that the SAR algorithm performs much better for highly correlated data sets than the other sparse CCA techniques, that impose prior covariance restrictions. This might explain why the SAR algorithm outperforms its alternatives in the out-of-sample cross-validation exercise. 

\begin{figure}
\begin{center}
\includegraphics[width=15cm]{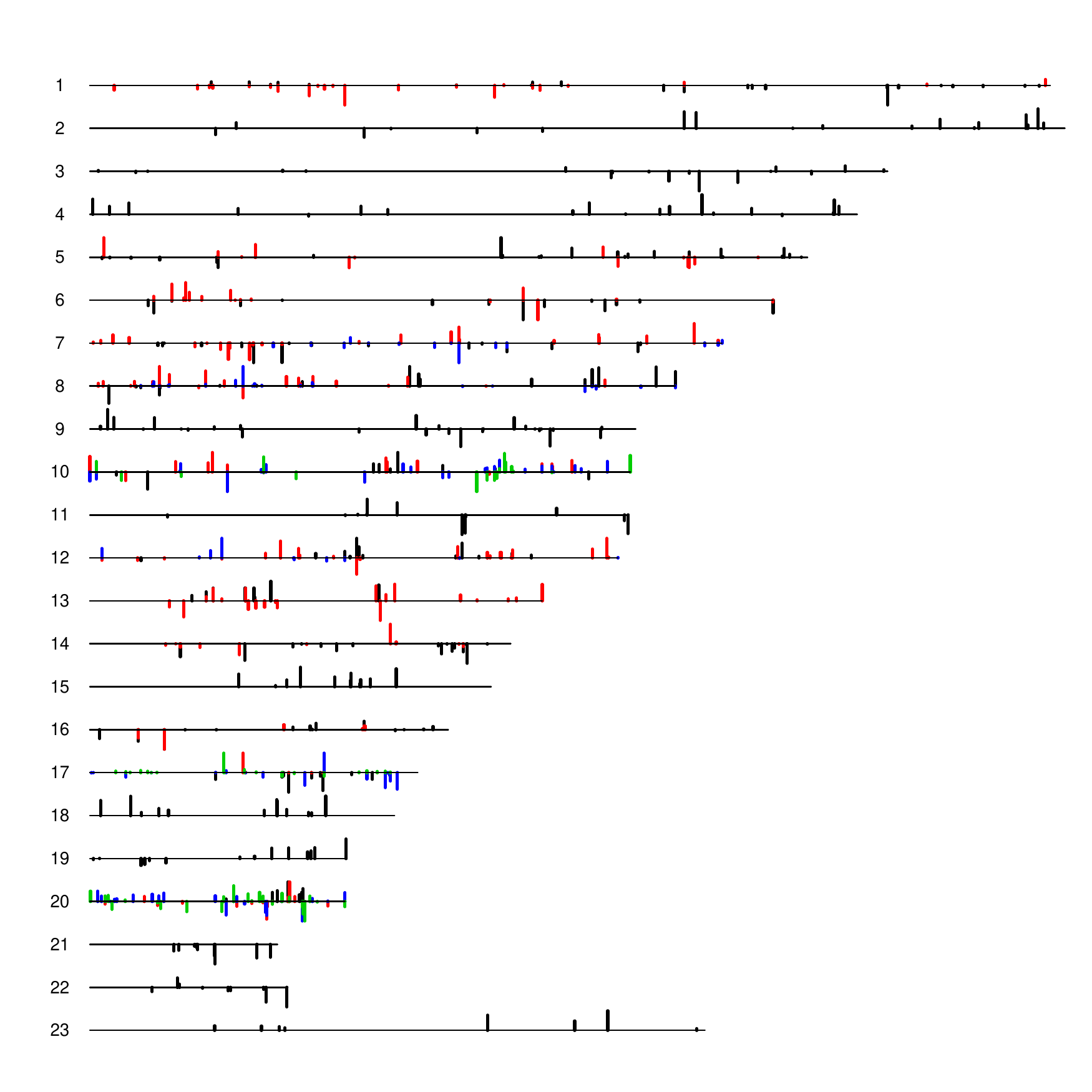} 
\caption{SAR algorithm: copy number change measurements with non-zero weights in the first (black), the second (red), the third (blue) and the fourth (green) canonical vectors are indicated for each of the 23 chromosomes.} \label{SMARapplication}
\end{center}
\end{figure}

Next, we discuss the solution provided by the SAR algorithm. For each chromosome, sparse canonical vectors are obtained. We do not fix the number of non-zero elements in the canonical vectors in advance, but select the sparsity parameter using the BIC discussed in Section 3. Figure \ref{SMARapplication} represents for each chromosome the copy number change measurements with non-zero weights\footnote{The construction of this figure is similar to the one presented in \cite{Witten09}. We use the R-code available in the R package \texttt{PMA} \citep{Rpma}.}. Each CGH spot has a certain position on a chromosome, called the nucleotide position. The CGH measurements selected by the SAR algorithm are indicated by plotting a vertical line on their respective nucleotide position. The different colors indicate the subset of variables selected in the construction of the corresponding canonical variate pair (first pair: black, second: red, third: blue, fourth: green).

We see from Figure \ref{SMARapplication} that the degree of sparsity selected by the BIC varies from one chromosome to the other. For chromosome 15, for example, only one canonical variate pair is selected and the BIC suggests a very sparse canonical vector. For chromosome 1, two canonical variate pairs are extracted with a large number of non-zero elements in the second canonical variate pair. However, a lot of non-zero weights are small in magnitude which can be seen from the length of the vertical lines. By adjusting the sparsity parameter to a higher value, a sparser solution could be obtained. A trade-off needs to be made between inducing more sparsity and thus performing better noise filtering, on the one hand, and reducing the risk of not including all important variables, on the other hand. Depending on the researcher's objective, the desired level of sparsity can be easily controlled by adjusting the sparsity parameter.

\section{Conclusion}
In high-dimensional settings, the estimation imprecision of traditional CCA will be large. To overcome this problem, penalized versions of CCA have been introduced such as the canonical ridge or sparse CCA. The canonical ridge still includes all variables in the canonical vectors, whereas sparse CCA only includes a subset of the variables. This is highly valuable in high-dimensional settings since it eases interpretation, as illustrated in the genomic data application.

In this paper, we introduce a Sparse Alternating Regression (SAR) algorithm that considers the CCA problem from a predictive point of view. We recast the CCA problem into a penalized alternating regression framework to obtain sparse canonical vectors. Contrary to other popular sparse CCA procedures (i.e. \citealp{Witten09}; \citealp{Parkhomenko09}; \citealp{Waaijenborg08}) we do not impose any covariance restrictions. We show that the SAR algorithm produces much better results than the other sparse CCA approaches. Especially in simulation settings when there is a dependency structure within each set of variables, the gains in estimation accuracy achieved by the SAR algorithm are outspoken. Also in the genomic data application, the data sets contain highly correlated variables. We illustrate that the SAR algorithm considerably outperforms the other sparse CCA techniques in an out-of-sample cross-validation exercise.

Both the SAR algorithm and the method of \cite{Waaijenborg08} use an alternating regression framework. There are, however, two important differences between both approaches, leading towards significant differences in performance. First, \cite{Waaijenborg08} perform univariate soft thresholding, which ignores the dependency structure within each set of variables. In contrast, we apply the lasso penalty to multiple linear regressions. The lasso only equals the soft thresholding estimator for a linear model with orthonormal design (see e.g. \citealp{Donoho94}). Secondly, we express the higher order canonical vectors in terms of the original data sets, whereas \cite{Waaijenborg08} express them in terms of the deflated data matrices. %We show in the simulation study and the genomic data example that the SAR algorithm performs significantly better.

In this paper, a lasso penalty is used to induce sparsity. Future work might consider other choices of penalty functions (see \citealp{Prabhakar2012}). For instance, the adaptive lasso \citep{Zou06}, the smoothly clipped absolute deviation (SCAD) penalty \citep{Fan01}, or a lasso with positivity constraints (see \citealp{Lykou10}). Note that \cite{Lykou10} also treat CCA as a least squares problem. They focus on orthogonality properties of CCA and only construct the first two pairs of sparse canonical vectors. Their approach could be extended to higher order canonical correlations, but this would increase the number of orthogonality constraints and the computing time substantially.

The level of sparsity produced by all sparse CCA techniques hinges on the selection method used for the sparsity parameters. This might lead to substantial differences in sparsity recognition performance, as illustrated in the simulation study. Future work still needs to be done on the comparison of methods (BIC, cross-validation, measure of explained variability, among others) to select the optimal tuning parameters.

\spacingset{1.2}
\bibliographystyle{asa}
\bibliography{sparsecancor}

\end{document}